\documentclass[a4paper,11pt]{article}

\pdfoutput=1 

\usepackage{jcappub} 

\usepackage[T1]{fontenc} 

\title{Can we hear beats with pulsar timing arrays?}

\author[a]{Shun Yamamoto}
\author[a]{Hideki Asada} 

\affiliation[a]
{Graduate School of Science and Technology, 
Hirosaki University,
Hirosaki 036-8561, Japan}

\emailAdd{yamamoto@tap.st.hirosaki-u.ac.jp}
\emailAdd{asada@hirosaki-u.ac.jp}

\abstract
{An isolated supermassive black hole binary (SMBHB) 
produces an identical cross-correlation pattern of pulsar timings 
as an isotropic stochastic background gravitational waves (GWs) 
generated possibly by inflation. 
Can there remain the identical cross-correlation pattern 
in the presence of a secondary SMBHB? 
To address this issue, 
the present paper focuses on 
GWs with similar amplitudes but slightly different frequencies 
$f_1$ and $f_2$ coming from two different directions.  
Beats between the two GWs 
can modify angular correlation patterns. 
The beat-induced correlation patterns 
are not stationary but modulated with a beat frequency 
$f_{beat} \equiv |f_1 - f_2|$. 
We obtain an analytic solution that allows us 
to infer $f_{beat}$ 
from the modulated angular correlations.}

\arxivnumber{2501.13450}

\begin{document}
\maketitle
\flushbottom

\section{Introduction}
The idea of using radio pulse timing to search for gravitational waves (GWs) 
can be dated back to Reference 
\cite{Estabrook, Sazhin, Detweiler}.
As a pioneering work, 
Hellings and Downs (HD) 
found that the sky correlation pattern among pulsar pairs, 
which depends upon the angle $\gamma$ 
between the lines of sight to the pulsars viewed from Earth, 
can be a strong evidence of GWs 
\cite{Hellings}, 
because it reflects the quadrupole nature of GWs 
\cite{CreightonBook, MaggioreBook, Anholm2009, Jenet}. 
See e.g. Reference 
\cite{Romano,Romano2024} 
for a review on detection methods of stochastic GW backgrounds. 
Eventually, 
several teams of pulsar timing arrays (PTAs) have recently 
reported a strong evidence of nano-hertz GWs 
\cite{Agazie2023, Antoniadis2023, Reardon2023, Xu2023}. 
According to their papers, 
superpositions of supermassive black hole binaries (SMBHBs)
are among possible GW sources, 
though improved methods, other possible explanations, and new physics 
have been vastly pursued 
e.g. \cite{Chen2020, Ellis, Smarra, Gouttenoire, Franciolini2023, DeRocco, Franciolini2024, Figueroa, Athron, Shih, Kumar, Bian, Chen2024}. 

For an isotropic stochastic background of GWs 
composed of the plus and cross polarization modes 
in general relativity, 
the expected correlated response of pulsar pairs follows the HD curve. 
It is worthwhile to mention that 
an isolated SMBHB 
produces an identical cross-correlation pattern 
as an isotropic stochastic background 
\cite{Cornish}.

Can there remain the identical cross-correlation pattern 
in the presence of a secondary SMBHB? 
In order to address this issue, 
the present paper considers 
GWs with slightly different frequencies $f_1$ and $f_2$ coming from two different directions. 
For a model as a confusion-noise case studied 
in References \cite{Allen2023,Allen2024}, 
all sources have an identical fixed GW frequency, 
such that they all lie in the same frequency bin. 
The resultant correlation pattern is stationary 
and identical as the HD curve.

\begin{figure}
\includegraphics[width=15.0cm]{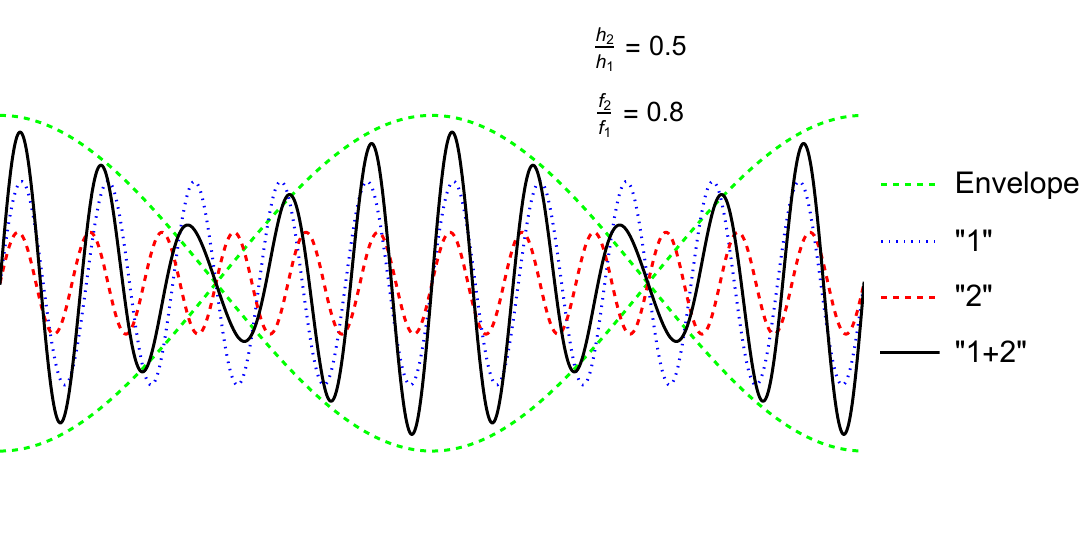}
\caption{
Beat between two sinusoidal waves traveling along the same line. 
$h_1$ and $f_1$ are the amplitude and frequency of the primary wave, 
while $h_2$ and $f_2$ are those of the secondary one. 
Here, $h_2/ h_1 = 0.5$ and $f_2/f_1 = 0.8$ are chosen. 
The blue, red and black (in color) curves denote the primary, secondary, 
and sum of sinusoidal waves and  respectively, 
where a green curve means an envelope wave of frequency $(f_1 - f_2)/2$. 
}
\label{fig-beat}
\end{figure}

Yet, 
one may ask if 
the situation can be changed 
when GWs with slightly different frequencies $f_1$ and $f_2$
($f_1 \approx f_2$) are considered. 
This question arises because an interference (called a beat) between the two GWs 
can increase (or decrease) the superimposed GW signal. 
The superimposed GW signal is not stationary but modulated. 
Figure \ref{fig-beat} shows an example of beats between two sinusoidal waves.

The main purpose of this paper is 
to examine if the beat frequency $f_{beat} = |f_1 - f_2|$ can be obtained 
in principle from modulated correlation patterns. 
In order to lay a theoretical groundwork for this issue, 
the present paper focuses on an ideal setup of 
two GWs with similar amplitudes but slightly different frequencies 
coming from two different directions, 
though 
it is much more likely that a few dozen binaries dominate
the PTA signal and numerical approaches are more preferred 
for such numerous binaries. 
If one GW amplitude is much larger, 
the PTA signal can fit well with the significantly largest GW component 
whereas much smaller GW components make small perturbations 
that are not distinguishable from noises, 
for which beats would be too small to be observed. 
Even so, the present groundwork can potentially have a practical application, 
if two dominant GWs in a certain frequency bin happen to have 
comparable amplitudes with similar frequencies as a rare case.

This paper is organized as follows. 
Section II discusses a modulated angular correlation pattern 
due to an interference between GWs 
with slightly different frequencies coming from two sky locations. 
Section III demonstrates that the beat frequency $f_{beat}$ 
can be measured in principle 
from the modulated correlations. 
Section VI is devoted to Summary.
Throughout this paper, 
the unit of $c=1$ is used and 
the center of the coordinates is the solar barycenter, 
safely approximated as the Earth. 
The superscripts $I =1$ and $2$ refer to the primary GW and secondary one, 
while 
the subscripts $K = 1, 2, 3$ and $4$ refer to four equal time domains 
that are defined in Section II.

\section{Modulating PTA angular correlations}
\subsection{Basic formulation} 
"Source averaging" is a useful calculational method 
for isotropic GW sources in PTA 
e.g. \cite{CreightonBook, MaggioreBook, Anholm2009, Jenet}. 
On the other hand, 
"pulsar averaging" was first introduced in Cornish and Sesana (2013) 
\cite{Cornish}, where an isolated SMBH is focused on. 
In the present paper, we use "pulsar averaging" 
for monochromatic and planar gravitational waves 
coming from two sky locations. 
See References \cite{McGrath2021,McGrath2022,Guo,Kubo,D'Orazio} 
for possible corrections by a nearby SMBHB and their cosmological implication. 
In addition, noise terms are ignored, 
because the present paper focuses on a proof-of-principle.

We consider two compact GW sources represented by SMBHBs,  
labeled by $I$ $(I =1, 2)$, 
for which $f_I$, $\hat\Omega_I$ and $h_{IA}$ 
for $A = +, \times$ as the plus and cross modes denote 
the intrinsic frequency of GWs,  
the source direction with respect to the Earth, 
and the amplitude of the $I$-th GW, respectively. 
The separation angle between the two GW directions is 
$\beta$, defined by 
$\cos \beta = \hat\Omega_1 \cdot \hat\Omega_2$ 
\cite{Allen2023,Allen2024}.

The waveform at the Earth is a superposition of GWs from the two sources, 
which can be expressed as 
\cite{CreightonBook, MaggioreBook}
\begin{align}
h_{i j}(t) 
=
\sum_{I=1, 2} 
\sum_{A= +,\times}  
h_{IA}
\exp [i \varphi_I(t)]  e_{i j}^A(\hat\Omega_I) ,
\label{h}
\end{align} 
where the phase $\varphi_I(t)$ is a periodic function in time. 
We follow also Reference \cite{Anholm2009} 
to adopt the Fourier representation in a complex function. 
We focus on a circular motion of a SMBHB for its simplicity, 
such that the phase can be written as $\varphi_I(t) = 2 \pi f_I (t + t_I)$.  
Here, 
$t_I$ is a time defined by $\Phi_I$
denoting the initial phase of the $I$-th GW 
($\Phi_I \equiv 2\pi f_I t_I$), 
and 
$e_{i j}^A(\hat\Omega_I)$ denotes the polarization tensor along the direction $\hat\Omega_I$. 
See e.g. Reference \cite{MaggioreBook, Poisson} 
for $\varphi_I(t)$ in elliptic motions.

We focus only on the Earth terms because pulsar terms vanish 
in the full-sky average 
\cite{Hellings, Cornish, Romano2024}. 
At the Earth, 
the redshift of radio signals from the $a$-th pulsar 
due to GWs from the two directions 
can be written as 
\cite{Anholm2009}
\begin{equation}
z_a(t) = \sum_{I = 1, 2} z_{a I}(t) ,
\label{z-two}
\end{equation}
where 
the contribution from the $I$-th direction is denoted as 
\begin{align}
z_{a I}(t) 
=&
\sum_{A= +,\times}  
F_a^A(\hat\Omega_I)
h_{IA}
\exp [i 2 \pi f_I (t + t_I)] , 
\label{z}
\end{align} 
and  
the antenna pattern function in PTAs for a pulsar in the direction $\hat{p}_a$ is 
\begin{align}
F_a^A(\hat\Omega_I)
=\frac{1}{2} \frac{\hat{p}_a^i \hat{p}_a^j}{1+\hat\Omega_I \cdot \hat{p}_a} e_{i j}^A(\hat\Omega_I) .
\label{F}
\end{align}

\subsection{Factorization of PTA angular correlations}
A full-sky averaged correlation between the $a$-th and $b$-th pulsars 
(in the directions $\hat{p}_a$ and $\hat{p}_b$, respectively)
with the separation angle $\gamma$
over the observation duration $T_{obs}$ is 
\begin{align}
\langle z_a z_b \rangle (\gamma) 
=&
\frac{1}{(4\pi)^2 T_{obs}} 
\int_{S^2} d\Omega_a 
\int_{S^2} d\Omega_b 
\int_0^{T_{obs}} dt 
\:
\delta(\cos \gamma - \hat{p}_a \cdot \hat{p}_b)
[z_a (t) z_b^*(t)] , 
\label{zz}
\end{align}
where $S^2$ means the 2-sphere centered at Earth, 
$\int_{S^2} d\Omega_{a}$ and $\int_{S^2} d\Omega_{b}$ 
denote the integration of the $a$-th and $b$-th pulsars over the full sky, respectively,  
the asterisk denotes the complex conjugate,  
and the Dirac's delta function is inserted to respect the separation angle 
\cite{Cornish}. 
In the presence of a secondary GW, 
Eq. (\ref{zz}) is the sum as 
$\langle z_a z_b \rangle 
= \langle z_{a 1} z_{b 1} \rangle + \langle z_{a 2} z_{b 2} \rangle 
+ \langle z_{a 1} z_{b 2} \rangle + \langle z_{a 2} z_{b 1} \rangle$. 
A diagonal term such as $\langle z_{a 1} z_{b 1} \rangle$ and $\langle z_{a 2} z_{b 2} \rangle$
is identical as Eq. (18) in Reference \cite{Cornish}.

\begin{figure}
\includegraphics[width=15.0cm]{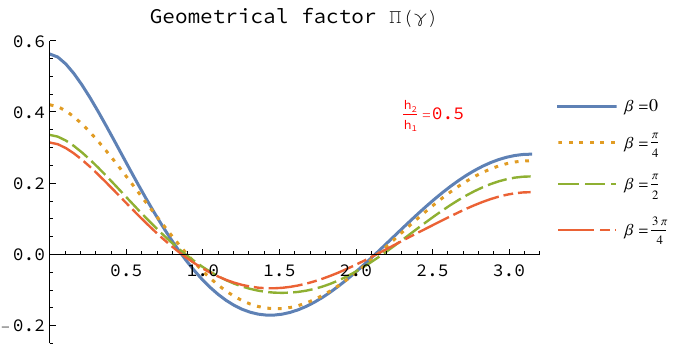}
\caption{
Geometrical factor $\Pi(\gamma)$ for $\beta = 0, \pi/4, \pi/2, 3\pi/4, \pi$, 
where 
$h_{1+} = h_{1\times}=1$ and $h_2 \equiv h_{2+} = h_{2\times}=0.5$ 
are chosen for its simplicity. 
See e.g. \cite{Cornish, Allen2024, Sasaki} 
for a set of coordinates convenient for numerical calculations of 
$\int_{S^2} d\Omega_{pa}$ and $\int_{S^2} d\Omega_{pb}$. 
}
\label{fig-Pi}
\end{figure}

Nontrivial contributions come from cross terms, 
$\langle z_{a 1} z_{b 2} \rangle + \langle z_{a 2} z_{b 1} \rangle$. 
For the later convenience, 
we consider a time domain as $t \in [t_i, t_f)$. 
The cross terms are factorized as 
\begin{align}
(\left\langle z_{a 1} z_{b 2}\right\rangle 
+ \left\langle z_{a 2} z_{b 1}\right\rangle) 
(\gamma; t_i, t_f) 
= 
\Pi(\gamma) 
\times 
B(t_i, t_f) ,
\label{crossterm}
\end{align}
where 
\begin{align}
\Pi(\gamma) 
=& 
\sum_{A, B= +,\times}  
\frac{h_{1 A} h_{2 B}}{(4\pi)^2} 
\int_{S^2} d\Omega_{a} 
\int_{S^2} d\Omega_{b} 
\:
F_a^A(\hat\Omega_1) F_b^B(\hat\Omega_2) 
\delta(\cos \gamma - \hat{p}_a \cdot \hat{p}_b) ,
\label{Pi}
\end{align}
\begin{align}
B(t_i, t_f) 
= 
\frac{1}{(t_f - t_i)} 
&\left[ 
\exp(i \Phi) \int_{t_i}^{t_f} dt \exp (i \omega_B t) 
+ \exp(-i \Phi) \int_{t_i}^{t_f} dt \exp (-i \omega_B t) 
\right] , 
\label{B}
\end{align}
for $\omega_B \equiv 2\pi (f_2 - f_1)$ denoting the beat angular frequency,
$\Phi \equiv \Phi_2 - \Phi_1$ denoting the initial phase difference of GWs.

$\Pi(\gamma)$ depends on a geometrical configuration 
characterized by 
the two GW directions ($\beta$) and  
and the pulsar pair separation ($\gamma$), 
while 
$B(t_i, t_f)$ reflects beat effects. 
Therefore, this paper refers to $\Pi(\gamma) $ and $B(t_i, t_f)$ 
as a geometrical factor and a beat factor, respectively. 
Note that the dependence of $\Pi(\gamma)$ on $\gamma$ 
is different from that in the HD angular correlation, 
because
the HD one is proportional to 
the diagonal term  
$\langle z_{a 1} z_{b 1} \rangle$ (and also $\langle z_{a 2} z_{b 2} \rangle$). 
See Figure \ref{fig-Pi} for numerical plots of $\Pi(\gamma)$.

After performing the time integration in Eq. (\ref{B}), 
the beat factor is 
\begin{align}
B(t_i, t_f) 
= 
\frac{2}{\omega_B (t_f - t_i)} 
\Bigl(
&
\cos \Phi 
[\sin(\omega_B t_f) - \sin(\omega_B t_i)] 
+ \sin\Phi
[\cos(\omega_B t_f) - \cos(\omega_B t_i)] \Bigr) . 
\label{B2} 
\end{align}
$B(t_i, t_f) 
\to 2 \cos\Phi$ 
in the noninterference limit ($f_1 = f_2$) 
in e.g. \cite{Allen2023,Allen2024}. 


\section{Can we hear beats?}
\subsection{Extracting a beat factor} 
A key is that 
the cross term $\langle z_{a 1} z_{b 2} \rangle + \langle z_{a 2} z_{b 1} \rangle$ 
is modulated via beat factors, 
while the diagonal term 
$\langle z_{a 1} z_{b 1} \rangle$ and $\langle z_{a 2} z_{b 2} \rangle$ are stationary. 
In order to extract the modulation, 
let us break the whole observation domain $T_{obs}$  
into four equal time intervals as 
$D_1 = \{t \in [0, T_{obs}/4)\}$, 
$D_2 = \{t \in [T_{obs}/4, T_{obs}/2)\}$, 
$D_3 = \{t \in [T_{obs}/2, 3T_{obs}/4)\}$, 
and 
$D_4 = \{t \in [3T_{obs}/4, T_{obs})\}$. 

From the above discussion, 
$\langle z_{a 1} z_{b 1} \rangle + \langle z_{a 2} z_{b 2} \rangle$ 
remains the same for any $D_K$  $(K =1, 2, 3$ and  $4)$, 
while $\langle z_{a 1} z_{b 2} \rangle + \langle z_{a 2} z_{b 1} \rangle$ 
depends upon $D_K$. 
The dependence of $B$ on $D_K$ 
plays a crucial role in the following calculations. 
In order to manifest this, $B_K$ denotes $B$ on $D_K$. 
Namely, $B_K = B((K-1)T_{obs}/4, KT_{obs}/4)$. 
The angular correlation over $D_K$ 
is denoted as  $\langle z_a z_b \rangle_K(\gamma)$.

\begin{figure}
\includegraphics[width=15.0cm]{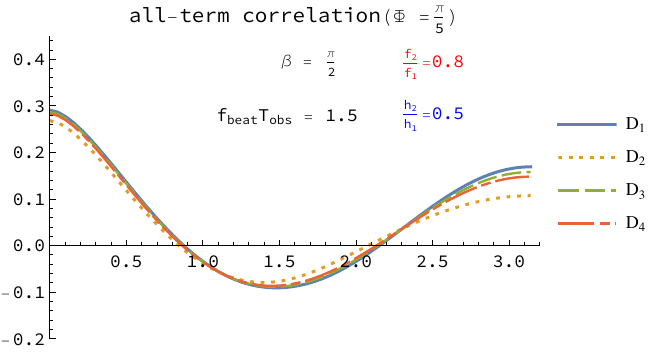}
\caption{
Angular correlations for GWs with $\beta = \pi/2$. 
For the curves to be recognized by eye, 
$h_2/ h_1 = 0.5$, $f_2/f_1 = 0.8$ and $\Phi = \pi/5$ 
are chosen as exaggerations. 
For the convenience in numerical plots, 
$h_1 \equiv h_{1+} = h_{1\times}$ and $h_2 \equiv h_{2+} = h_{2\times}$ 
are assumed, 
which affect only the magnitude of $\Pi(\gamma)$ but 
do not influence our discussions especially including Figure \ref{fig-B}. 
}
\label{fig-correlation}
\end{figure}

In order to quantify the differences among the $D_K$-dependent correlations, 
we define 
\begin{align}
\delta_{123} 
\equiv 
\frac{\langle z_a z_b \rangle_1(\gamma) - \langle z_a z_b \rangle_2(\gamma)}
{\langle z_a z_b \rangle_2(\gamma) - \langle z_a z_b \rangle_3(\gamma)} , 
\label{delta}
\end{align}
Cyclically, $\delta_{234}$ is defined, 
where $\delta_{123}$ and $\delta_{234}$ are defined for the same $\gamma$. 
One may thus ask if they depend on $\gamma$. 
We shall show below that they do not. 

The diagonal terms remain the same for any $D_K$, 
which leads to $\langle z_{a I} z_{b I} \rangle_K - \langle z_{a I} z_{b I} \rangle_L = 0$ 
for $K, L=1, \cdots, 4$. 
The diagonal terms thus vanish in $\delta_{123}$ and $\delta_{234}$. 
Therefore, only the cross terms contribute to 
$\delta_{123}$ and $\delta_{234}$ as 
\begin{align}
\delta_{123} 
=
\frac{
(\left\langle z_{a 1} z_{b 2}\right\rangle 
+ \left\langle z_{a 2} z_{b 1}\right\rangle)_1 
(\gamma) 
 - 
(\left\langle z_{a 1} z_{b 2}\right\rangle 
+ \left\langle z_{a 2} z_{b 1}\right\rangle)_2 
(\gamma)
}
{
(\left\langle z_{a 1} z_{b 2}\right\rangle 
+ \left\langle z_{a 2} z_{b 1}\right\rangle)_2 
(\gamma)
- 
(\left\langle z_{a 1} z_{b 2}\right\rangle 
+ \left\langle z_{a 2} z_{b 1}\right\rangle)_3 
(\gamma)
} . 
\label{delta2}
\end{align}

By using Eq. (\ref{crossterm}) in Eq. (\ref{delta2}), 
we obtain 
\begin{align}
\delta_{123} 
&=
\frac{B_1 - B_2}{B_2 - B_3} , 
\notag\\
\delta_{234} 
&=
\frac{B_2 - B_3}{B_3 - B_4}  , 
\label{delta3}
\end{align}
where we use that 
$(\left\langle z_{a 1} z_{b 2}\right\rangle 
+ \left\langle z_{a 2} z_{b 1}\right\rangle)_1, 
\cdots, 
(\left\langle z_{a 1} z_{b 2}\right\rangle 
+ \left\langle z_{a 2} z_{b 1}\right\rangle)_4$ 
have a common factor $\Pi(\gamma)$. 

Eq. (\ref{delta3}) means that 
$\delta_{123}$ and $\delta_{234}$ do not depend on $\gamma$. 
We can choose any value of $\gamma$ in principle, 
when $\delta_{123}$ and $\delta_{234}$ are evaluated from angular correlations. 
Note that 
$\delta_{123}$ and $\delta_{234}$ are free from $\beta$ too.

\begin{figure}
\includegraphics[width=15.0cm]{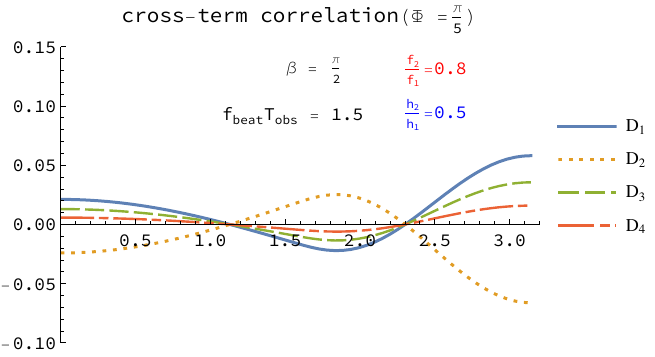}
\caption{
Cross-term correlation $\Pi(\gamma) B_K$ by 
Eq. (\ref{crossterm}) 
dependent on $D_K$. 
The parameters correspond to those in Figure \ref{fig-correlation}.} 
\label{fig-B}
\end{figure}

\subsection{Inverse problem and its solution}
The prefactor $2/\omega_B (t_f - t_i)$ in Eq. (\ref{B2}) cancels out 
in a fractional form as Eq. (\ref{delta3}). 
Eq. (\ref{delta3}) is thus rewritten as 
\begin{align}
&C_{123} \cos\Phi + S_{123} \sin\Phi = 0 , 
\notag\\
&C_{234} \cos\Phi + S_{234} \sin\Phi = 0 , 
\label{eqs}
\end{align}
where 
$\eta \equiv \omega_B T_{obs}/4$ (a phase of the beat during each time interval), and 
\begin{align}
C_{123} 
&\equiv 
\delta_{123} \sin3\eta - (1 + 2 \delta_{123}) \sin2\eta + (2 + \delta_{123}) \sin\eta ,
\notag\\
S_{123} 
&\equiv 
\delta_{123} \cos3\eta - (1 + 2 \delta_{123}) \cos2\eta + (2 + \delta_{123}) \cos\eta ,
\notag\\
C_{234} 
&\equiv 
\delta_{234} \sin4\eta - (1 + 2 \delta_{234}) \sin3\eta + (2 + \delta_{234}) \sin2\eta ,
\notag\\
S_{234} 
&\equiv 
\delta_{234} \cos4\eta - (1 + 2 \delta_{234}) \cos3\eta + (2 + \delta_{234}) \cos2\eta .
\label{CS}
\end{align}

Eq. (\ref{eqs}) can be rearranged as 
\begin{align}
\begin{pmatrix}
C_{123} &  S_{123}\\
C_{234} & S_{234} 
\end{pmatrix}
\begin{pmatrix}
\cos\Phi \\
\sin\Phi
\end{pmatrix}
= 
\begin{pmatrix}
0 \\
0
\end{pmatrix} .
\label{matrixform}
\end{align}
By using $(\cos\Phi, \sin\Phi) \neq (0, 0)$ in Eq. (\ref{matrixform}),  
we obtain 
\begin{align}
\begin{vmatrix}
C_{123} &  S_{123}\\
C_{234} & S_{234} 
\end{vmatrix}
= 0 . 
\label{det} 
\end{align}
This is an equation for $\eta$. 
By direct calculations, 
Eq. (\ref{det}) is factorized as 
\begin{align}
\sin\eta (\cos\eta - 1)^2 
[2 \delta_{234} \cos\eta - (1 + \delta_{123}\delta_{234})] 
= 0 ,
\label{det2}
\end{align}
where Eq. (\ref{CS}) is used. 

First, we consider a noninterference case ($f_1 = f_2$) 
(e.g. \cite{Allen2023, Allen2024}), 
which leads to $\eta = 0$. 
Namely, $\sin\eta = 0$ and $\cos \eta =1$. 
Hence, the noninterference case can be expressed by Eq. (\ref{det2}). 

Secondly, we study an interference case ($f_1 \neq f_2$), 
for which usually  $\sin\eta \neq 0$ ($\cos \eta \neq1$).  
Eq. (\ref{det2}) becomes $2 \delta_{234} \cos\eta - (1 + \delta_{123}\delta_{234})=0$. 
Therefore, $\eta$ is determined by  
\begin{align}
\cos\eta = \frac{1 + \delta_{123}\delta_{234}}{2 \delta_{234}} . 
\label{sol}
\end{align}
The beat frequency can be thus inferred in principle 
from modulated angular correlations. 
$\eta$ does not depend on the initial phase difference $\Phi$. 
This implies that the right-hand side of Eq. (\ref{sol}) does not depend on 
$\Phi$, though $\delta_{123}$ and $\delta_{234}$ do. 

Thirdly, we mention that interferences ($f_1 \neq f_2$) can have 
a particular case ($\sin\eta = 0$), 
for which Eq. (\ref{sol}) does not make sense. 
This particular case occurs when $(f_2 - f_1) T_{obs} = 2n$ for an integer $n$. 
Clearly, this case is very rare in the sense that 
$T_{obs}$ must happen to coincide with the integer multiple 
of the inverse of the beat frequency. 
This third case migrates to the second case 
when a different observational duration ($\neq T_{obs}$) is chosen.

\subsection{Numerical example}
Figure \ref{fig-correlation} shows a numerical example 
for angular correlations for four time domains $D_1, \cdots D_4$. 
Without loss of generality, 
$\hat\Omega_1$ and $\hat\Omega_2$ are located on the $x - z$ plane, 
such that the combination of the antenna pattern functions 
takes a diagonal form of 
$F_a^+(\hat\Omega_1) F_b^+(\hat\Omega_2) 
+ F_a^{\times}(\hat\Omega_1) F_b^{\times}(\hat\Omega_2)$. 
See also Appendix G of \cite{Allen2023} 
for the diagonal form.

Figure \ref{fig-B} shows beat factors $B_K$, 
which are proportional to the cross-term correlations, 
for the same parameters as those in Figure \ref{fig-correlation}. 

For the numerical example in Figures \ref{fig-correlation} and \ref{fig-B}, 
$\delta_{123} = -1.22085$ and $\delta_{234} = -5.17164$. 
By substituting these numbers into Eq. (\ref{sol}), 
we obtain $\eta = 2.356197$. 
This agrees with our numerical input $f_{beat} T_{obs} = 1.5$ 
with sufficient accuracy.

Let us briefly mention a possible range of parameters. 
The magnitude of the cross terms are proportional to $h_1 \times h_2$, 
while the leading term of the correlations is 
$\langle z_{a 1} z_{b 1} \rangle$ proportional to $(h_1)^2$. 
The ratio of the cross terms to the leading term is 
roughly $O(h_2/h_1)$. 
When correlations are measured with e.g. 20 percent accuracy, 
the beat modulation will be marginally detectable if $h_2/h_1 > 0.2$. 

Next, we mention a possible beat frequency. 
The current results hold when $f_{beat} T_{obs} \sim O(1)$. 
When $f_{beat} T_{obs} \gg 1$, there are many beat oscillations in one $D_K$, 
and thus the information about the beats is smeared out. 
It is thus difficult to hear beats in this case.  
When $f_{beat} T_{obs} \ll 1$, on the other hand, 
only a small portion of the beat oscillation is included in $D_K$ 
and thus it is difficult to determine $f_{beat}$. 
In order to clarify the parameter ranges, 
a detailed and numerical study is needed. 
It is left for future.

\subsection{Validity of the monochromatic assumption} 
We clarify the validity of the main assumption. 
There may be two effects that can be relevant with the monochromatic assumption. 
First, the chirp in an inspiral phase of a SMBHB causes the GW frequency evolution. 
The GW frequency shift over $T_{obs}$  can be expressed as 
$\Delta f \sim \dot{f}|_{GW} \times T_{obs}$, 
where $\dot{f}|_{GW}$ means the time derivative of the GW frequency 
owing to the GW radiation. 

When the frequency shift of the primary GW is comparable to 
that of the secondary one,  
the beat frequency remains almost constant, 
for which the above results remain unchanged. 
We investigate another case 
that the frequency shift of the primary GW is 
much larger than that of the secondary one 
($|\Delta f_1| \gg |\Delta f_2|$). 
In this case, the shift of the beat frequency is 
$|\Delta f_{beat}| \sim |\Delta f_1|$. 
We shall evaluate the frequency shift below. 

In the quadrupole approximation of GWs \cite{MaggioreBook,Poisson}, 
$\dot{f_1}|_{GW} = (96 \pi^{8/3}/5) (f_1)^{11/3} (M_{ch1})^{5/3}$, 
where we use 
the energy loss rate due to GWs 
and the Kepler's third law for the orbital motion, 
and  $M_{ch1}$ denotes 
the chirp mass  of the primary SMBHB. 

The PTA measurement accuracy of the GW frequency 
is roughly the inverse of $T_{obs}$, 
which leads to 
the condition that 
the monochromatic assumption for the cross term 
remains valid 
as $\Delta f_{beat} < 1/T_{obs}$. 
By bringing together these results, 
this inequality can be rearranged as 
\begin{align}
M_{ch1} 
< 
10^{10} M_{\odot} 
\left(\frac{1 \;\mbox{year}^{-1}}{f_1}\right)^{11/5}
\left(\frac{10 \;\mbox{year}}{T_{obs}}\right)^{6/5} .
\label{validity-1}
\end{align}

Therefore, 
the monochromatic assumption is valid for 
normal SMBHBs ($M_{ch} < 10^{10} M_{\odot}$) 
except for a SMBHB in the heaviest class ($M_{ch} \geq 10^{10} M_{\odot}$).  
In the case that $|\Delta f_1| \ll |\Delta f_2|$, 
on the other hand, 
Eq. (\ref{validity-1}) is still applicable but with the chirp mass of 
the secondary SMBHB denoted as $M_{ch2}$. 

Next, we consider an elliptic motion of a SMBHB with the eccentricity $e_K$, 
for which corrections to a sinusoidal form occur 
and the GW frequency changes with time \cite{MaggioreBook, Poisson}. 
The magnitude of the frequency shift is $\Delta f \sim f e_K$. 
The monochromatic treatment remains valid 
if $\Delta f < 1/T_{obs}$. 
For typical PTA parameters, 
this inequality leads to 
\begin{align}
e_K < 0.1 
\left(\frac{1 \;\mbox{year}^{-1}}{f}\right)
\left(\frac{10 \;\mbox{year}}{T_{obs}}\right) .
\label{validity-2}
\end{align}

For a nearly circular orbit ($e_K < 0.1$), 
the monochromatic assumption is valid, 
while the assumption does not hold 
for a large eccentricity that 
may imply a younger SMBHB. 

See also \cite{Hu} 
for discussions on different frequencies originating from the same GW source 
in testing gravity theories.

\subsection{Two GW sources over the stochastic background}
Finally, we mention a possible way to make the scenario more realistic. 
It is to consider two nearby bright GW sources that dominate over the
stochastic background. 
In this case, 
the GW perturbation is the sum as $h_{1A} + h_{2A} + h_{BG A}$ for $A = +, \times$, 
where the subscript $BG$ denotes the stochastic background. 
The stochastic background has no correlation with the primary GW nor the secondary one. 
Hence, $<z_{a 1}  z_{b BG}> = 0$, $<z_{a 2}  z_{b BG}> = 0$, 
$<z_{a BG}  z_{b 1}> = 0$, and $<z_{a BG}  z_{b 2}> = 0$, 
where $z_{a BG}$ and $z_{b BG}$ denote 
the redshifts due to the stochastic background. 

On the other hand, 
the stochastic background makes correlations as 
$<z_{a BG}  z_{b BG}> \neq 0$, 
which is the usual HD correlation. 
As discussed already, 
the HD correlation is stationary to cancel out in Eq. (\ref{delta}). 
Therefore, Eqs.(\ref{delta2})-(\ref{sol}) 
remain unchanged. 
The results and discussions in the present paper 
are thus valid also in this scenario.

\section{Summary}
Beat effects between 
GWs with similar amplitudes but slightly different frequencies coming from two sky directions 
have been discussed. 
Angular correlations can be modulated 
by the beat effects. 
We have found an analytic solution that allows us 
to infer $f_{beat}$ 
from the modulated angular correlations.
Namely, in principle, we can hear beats with PTAs. 

Along the direction of this paper, 
it would be interesting to perform detailed calculations 
taking account of observational noises in order to make an estimation of 
a possible accuracy in measured $f_{beat}$. 
The present paper focuses on the minimum number of equal time intervals, 
namely four time domains. 
What is the optimal number of $D_K$ in a realistic case? 
This issue is left for future.

\section{Acknowledgments}
We would like to thank J. D. E. Creighton for useful comments 
on the earlier version of the manuscript. 
We are grateful to 
Keitaro Takahashi and Yuuiti Sendouda 
for useful discussions. 
We thank Tatsuya Sasaki, Kohei Yamauchi, Taichi Ueyama and Ryuya Kudo 
for fruitful conversations. 
We wish to thank JGRG33 workshop participants for stimulating conversations. 
This work was supported 
in part by Japan Society for the Promotion of Science (JSPS) 
Grant-in-Aid for Scientific Research, 
No. 20K03963 (H.A.) and 24K07009(H.A.).

\end{document}